\documentstyle[preprint,epsf,eqsecnum,aps,floats]{revtex}
\begin{document}
\title{Search for Right-Handed $W$ Bosons and Heavy $W^\prime$ in
  $p\bar p$ Collisions at $\sqrt{s} =$1.8 TeV}
%
\author{
S.~Abachi,$^{14}$
B.~Abbott,$^{28}$
M.~Abolins,$^{25}$
B.S.~Acharya,$^{44}$
I.~Adam,$^{12}$
D.L.~Adams,$^{37}$
M.~Adams,$^{17}$
S.~Ahn,$^{14}$
H.~Aihara,$^{22}$
J.~Alitti,$^{40}$
% ...
\\
\vskip 0.50cm
\centerline{(D\O\ Collaboration)}
\vskip 0.50cm
}
\address{
\centerline{$^{1}$Universidad de los Andes, Bogot\'{a}, Colombia}
\centerline{$^{2}$University of Arizona, Tucson, Arizona 85721}
\centerline{$^{3}$Boston University, Boston, Massachusetts 02215}
\centerline{$^{4}$Brookhaven National Laboratory, Upton, New York 11973}
\centerline{$^{5}$Brown University, Providence, Rhode Island 02912}
\centerline{$^{6}$Universidad de Buenos Aires, Buenos Aires, Argentina}
\centerline{$^{7}$University of California, Davis, California 95616}
\centerline{$^{8}$University of California, Irvine, California 92717}
\centerline{$^{9}$University of California, Riverside, California 92521}
\centerline{$^{10}$LAFEX, Centro Brasileiro de Pesquisas F{\'\i}sicas,
                  Rio de Janeiro, Brazil}
\centerline{$^{11}$CINVESTAV, Mexico City, Mexico}
\centerline{$^{12}$Columbia University, New York, New York 10027}
\centerline{$^{13}$Delhi University, Delhi, India 110007}
\centerline{$^{14}$Fermi National Accelerator Laboratory, Batavia,
                   Illinois 60510}
\centerline{$^{15}$Florida State University, Tallahassee, Florida 32306}
\centerline{$^{16}$University of Hawaii, Honolulu, Hawaii 96822}
\centerline{$^{17}$University of Illinois at Chicago, Chicago, Illinois 60607}
\centerline{$^{18}$Indiana University, Bloomington, Indiana 47405}
\centerline{$^{19}$Iowa State University, Ames, Iowa 50011}
\centerline{$^{20}$Korea University, Seoul, Korea}
\centerline{$^{21}$Kyungsung University, Pusan, Korea}
\centerline{$^{22}$Lawrence Berkeley National Laboratory and University of
                   California, Berkeley, California 94720}
\centerline{$^{23}$University of Maryland, College Park, Maryland 20742}
\centerline{$^{24}$University of Michigan, Ann Arbor, Michigan 48109}
\centerline{$^{25}$Michigan State University, East Lansing, Michigan 48824}
\centerline{$^{26}$Moscow State University, Moscow, Russia}
\centerline{$^{27}$University of Nebraska, Lincoln, Nebraska 68588}
\centerline{$^{28}$New York University, New York, New York 10003}
\centerline{$^{29}$Northeastern University, Boston, Massachusetts 02115}
\centerline{$^{30}$Northern Illinois University, DeKalb, Illinois 60115}
\centerline{$^{31}$Northwestern University, Evanston, Illinois 60208}
\centerline{$^{32}$University of Notre Dame, Notre Dame, Indiana 46556}
\centerline{$^{33}$University of Oklahoma, Norman, Oklahoma 73019}
\centerline{$^{34}$University of Panjab, Chandigarh 16-00-14, India}
\centerline{$^{35}$Institute for High Energy Physics, 142-284 Protvino, Russia}
\centerline{$^{36}$Purdue University, West Lafayette, Indiana 47907}
\centerline{$^{37}$Rice University, Houston, Texas 77251}
\centerline{$^{38}$Universidade Estadual do Rio de Janeiro, Brazil}
\centerline{$^{39}$University of Rochester, Rochester, New York 14627}
\centerline{$^{40}$CEA, DAPNIA/Service de Physique des Particules, CE-SACLAY,
                   France}
\centerline{$^{41}$Seoul National University, Seoul, Korea}
\centerline{$^{42}$State University of New York, Stony Brook, New York 11794}
\centerline{$^{43}$SSC Laboratory, Dallas, Texas 75237}
\centerline{$^{44}$Tata Institute of Fundamental Research,
                   Colaba, Bombay 400005, India}
\centerline{$^{45}$University of Texas, Arlington, Texas 76019}
\centerline{$^{46}$Texas A\&M University, College Station, Texas 77843}
}

\date{\today}
\maketitle

\begin{abstract}
\indent
We report on a search for right-handed $W$ bosons ($W_R$).
We used data collected with the D\O \ detector at the Fermilab Tevatron
$p\bar p$ collider at $\sqrt{s}=$1.8 TeV to search for $W_R$ decays
into an electron and a massive right-handed neutrino
$W_R^\pm \rightarrow  e^\pm N_R$.
Using the inclusive electron data,
we set mass limits independent of the $N_R$ decay:
$m_{W_R}>650$ GeV/c$^2$ and $m_{W_R}>720$ GeV/c$^2$ at the 95\% confidence
level,
valid for $m_{N_R}<\frac{1}{2}m_{W_R}$ and $m_{N_R} \ll m_{W_R}$ respectively.
The latter also represents a new lower limit on the mass of a heavy
left-handed $W$ boson ($W^\prime$) decaying into $e \nu$.
In addition, limits on $m_{W_R}$ valid for larger values of the $N_R$ mass
are obtained assuming that $N_R$ decays to an electron and two jets.
\end{abstract}

\indent
 Right-handed $W$ gauge bosons ($W_R$) are additional intermediate vector
particles that arise in extensions of the Standard Model (SM) such as the
left-right symmetric model (LRM) \cite{Unification}. In the LRM, an enlarged
$SU(2)_R \times SU(2)_L \times U(1)$ symmetry group replaces the
$SU(2)_L \times U(1)$ group of the SM. As a result of the additional symmetry,
three new gauge bosons, two charged $W_R^\pm$ and one neutral $Z'$, appear
along with massive right-handed neutrinos ($N_R$).

\indent
 In this Letter, a direct search for $W_R$ bosons with mass greater than 200
GeV/c$^2$ which decay into an electron (or positron)
and a massive right-handed neutrino, $W_R \rightarrow eN_R$
\cite{REF_WR_eN} is reported.
The $N_R$ is assumed to decay promptly through the
right-handed charged current into a mode that depends on the
mixing angle $\xi$ between $W_L$ and $W_R$.
If the mixing is negligible (no mixing case), the $N_R$ will decay
into an electron and an off-shell
$W_R$, $N_R \rightarrow eW_R^*$.
The right-handed neutrinos from other lepton families
are assumed to be at least as massive as the electron-$N_R$.
Therefore, the off-shell $W_R$ can decay only into quarks,
$W_R^* \rightarrow q\bar q'$.
On the other hand, if the mixing is large,
the $N_R$  decays into an electron and a $W$ boson, which decays into quarks
two thirds of the time.
In both cases the decay chain leads predominantly to a final state with
$eeqq$.

\indent
Previous direct searches at hadron colliders yielded the lower limits
$m_{W_R}>261$ GeV/c$^2$ \cite{UA2_DIJET}, valid for any value of the mass of
the right-handed neutrino, and $m_{W_R}>652$ GeV/c$^2$ \cite{CDF_Wprime},
valid only for a light right-handed neutrino ($m_{N_R} \ll m_{W_R}$)
that does not decay or interact within the detector.
Indirect searches based on low energy phenomena
such as $\mu$ decay, the $K_L$-$K_S$ mass difference, and neutrinoless
double beta decay provide  additional
stringent lower limits \cite{Existing_Limits}.
Limits from direct and indirect searches depend, however,
on the assumed values of the elements of the mixing matrix $V^R$ for the
right-handed quarks,
the coupling constant $g_R$, the mass and type (Dirac or Majorana)
of the right-handed neutrinos, and the
mixing angle $\xi$. The most general limit is
$m_{W_R} \cdot \frac{g_L}{g_R}>300$ GeV/c$^2$ \cite{Existing_Limits}.

\indent
 Two different methods, corresponding to different
values of the ratio $R_m=m_{N_R}$/$m_{W_R}$,
are used for this search. For $R_m
{\raise0.3ex\hbox{$\;<$\kern-0.75em\raise-1.1ex\hbox{$\sim\;$}}}
\frac{1}{2}$,
the products of the $N_R$ decay are not likely to be well separated,
making their individual identification difficult.
Therefore, the transverse momentum spectrum of the
$W_R$ decay electron, which is expected to be hard and to have a
distinctive Jacobian peak at $(m_{W_R}^2-m_{N_R}^2)/2m_{W_R}$,
is used as a signature.
A search for such a peak, henceforth referred to as the {\sl peak search},
is carried out using the high-$p_T$ inclusive electron data.
This method does not discriminate between helicities of the $W$ boson.
Therefore, the {\sl peak search} is also sensitive
to heavy left-handed $W$ bosons ($W^\prime$) which decay into an electron
and an electron neutrino $W^\prime \rightarrow e\nu$.
For $R_m
{\raise0.3ex\hbox{$\;>$\kern-0.75em\raise-1.1ex\hbox{$\sim\;$}}}
\frac{1}{2}$, the
products of the $N_R$ decay are likely to be well separated, making possible
the detection of the exclusive final state with two electrons and two jets.
After requiring the two electrons to be inconsistent with
$Z \rightarrow ee$ decay, the background due to other known physics
processes is small.
Therefore, a simple counting experiment, referred to here as the
{\sl eejj search}, is performed.
The analysis presented here is based on approximately 79 pb$^{-1}$ of data
collected during two Fermilab Tevatron $p\overline{p}$ collider runs at
$\sqrt{s}=$1.8 TeV from August 1992 to May 1995.

\indent
The  D\O\ detector consists of three major subsystems:
a central tracking system with no magnetic field, a hermetic
uranium-liquid argon sampling calorimeter, and a muon magnetic spectrometer.
The calorimeter has fine longitudinal and transverse segmentation in
pseudorapidity ($\eta$) and azimuth ($\phi$) that allows
electromagnetic showers to be distinguished from jets.
It provides full coverage for $|\eta|\leq$
4 with energy resolution 15\%/$\sqrt{E({\rm GeV})}$ for electromagnetic
showers and 80\%/$\sqrt{E({\rm GeV})}$ for hadronic jets.
The central and forward drift chambers are used to identify charged tracks for
$|\eta|\leq$ 3.1 and to locate the primary vertex. A more detailed description
of the D\O\ detector can be found elsewhere \cite{D0Detector}.

\indent
To identify electrons \cite{D0_TOP_PRD}, the presence of an isolated
electromagnetic energy cluster
with shape consistent with that of an electron
(as determined from test beam measurements) is required.
In addition, an associated charged track that matches the calorimeter
cluster in $\eta$ and $\phi$ and with an ionization in the drift chambers
consistent with that of a minimum ionizing particle must be found.
Jets are reconstructed using a cone algorithm with a cone radius of 0.5 in
$\eta$-$\phi$ space.

\indent
For the {\sl peak search}, events were collected using a
single electromagnetic cluster trigger.
Offline, the inclusive high-$p_T$ electron events
were selected by requiring an electron candidate
with $p_T^e>55$ GeV/c and $|\eta_e|<1.1$.
To reduce the multijet background (QCD) from events with a jet misidentified
as an electron,
strict electron identification criteria were imposed.
The 101 events with $p_T^e>100$ GeV/c were scanned to search for
anomalies; we discarded one event which was consistent with being a high
energy cosmic ray muon that showered in the electromagnetic part of the
calorimeter, mimicking an electron.

\begin{figure}[t]
\epsfxsize=3.375in
\centerline{\leavevmode\epsffile[0 0 567 370]{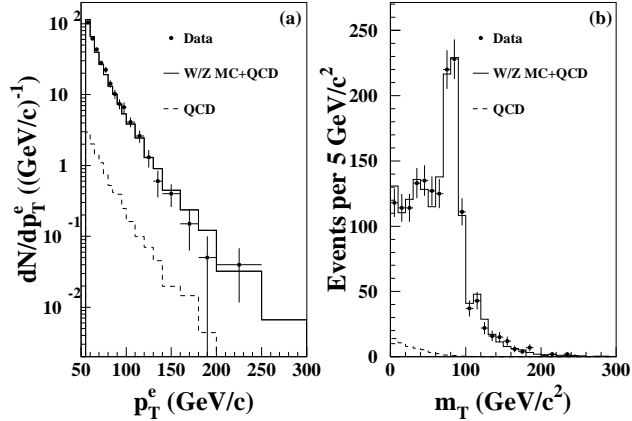}}
\vspace{0.3cm}
\caption{(a) Electron transverse momentum and (b) transverse mass
(formed by the electron and
\mbox{${\hbox{$E$\kern-0.6em\lower-.1ex\hbox{/}}}_T$}) distributions
of the inclusive high-$p_T$ electron sample.}
\label{PRL_FITS}
\end{figure}

\indent
The primary background in the {\sl peak search} is due to
highly off-shell and
large-$p_T$ $W$ and $Z$ boson production. These processes were simulated
using a Monte Carlo (MC) program based on
a theoretical calculation of the bosons' $p_T$ \cite{Arnold-Kaufman} and
on the bosons' line shape obtained using the {\footnotesize PYTHIA}
\cite{PYTHIA} MC program, with
a simple detector simulation. The QCD background was modeled using the
collider data.

\indent
A simultaneous fit to the transverse mass
($m_T$)
distribution,
formed by the electron and the missing transverse energy
\mbox{${\hbox{$E$\kern-0.6em\lower-.1ex\hbox{/}}}_T$}, and to the
electron transverse momentum ($p_T^e$) distribution was performed.
A binned maximum likelihood fit was used to find the contributions
of the combined $W$ and $Z$ boson backgrounds and the
QCD background \cite{My_thesis}.
Figure \ref{PRL_FITS} shows the $p_T^e$ and
$m_T$ distributions with their
corresponding fits.
The confidence level (CL) is 71\% for the  $p_T^e$ fit and 90\% for the
$m_T$ fit.
The presence of $W_R \rightarrow eN_R$ decays would appear as an excess in
a few consecutive bins in the $p_T^e$ distribution. No evidence for such an
excess is observed.

\indent
The acceptance and $p_T^e$ distribution of the signal were
obtained for a grid of points in the ($m_{W_R}$,$m_{N_R}$) plane
using {\footnotesize PYTHIA} MC samples
with a detector simulation based on the {\footnotesize GEANT}
program \cite{GEANT}.
The 95\% CL
upper limit on the number of
$W_R$ events was obtained by integrating the probability
\begin{figure}[h]
\epsfxsize=3.375in
\centerline{\leavevmode\epsffile[0 0 567 520]{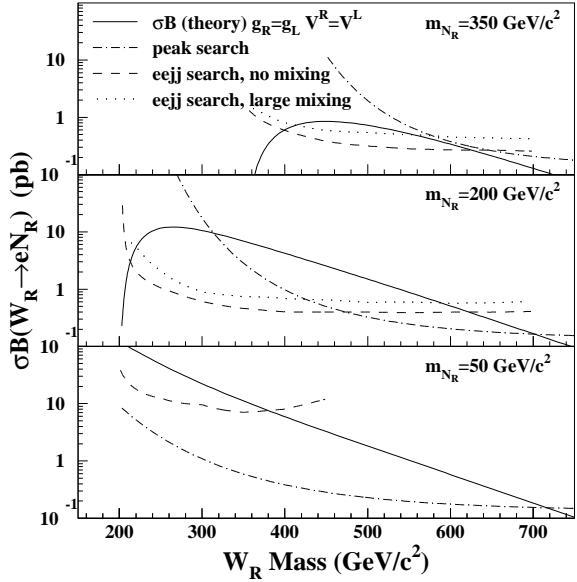}}
\vspace{0.3cm}
\caption{95\% CL upper limit on $\sigma B$ as a function
of the \protect \nolinebreak[4]
$W_R$ boson mass. Limits are shown for three
values of the $N_R$ mass.}
\label{PRL_LIMIT_SB_NEW}
\end{figure}
of the presence of a $W_R$ component in the measured $p_T^e$
distribution for every point in the grid.
This was converted into an upper limit on the cross
section times branching fraction ($\sigma B$) by normalizing to the
measured $W$ and $Z$ boson production cross sections \cite{D0_W_Cross}
using the observed $W/Z$ component in the initial simultaneous $p_T^e$
and $m_T$ fit and the acceptances as calculated from MC simulation.

\begin{figure}[t]
\epsfxsize=3.375in
\centerline{\leavevmode\epsffile[0 0 567 520]{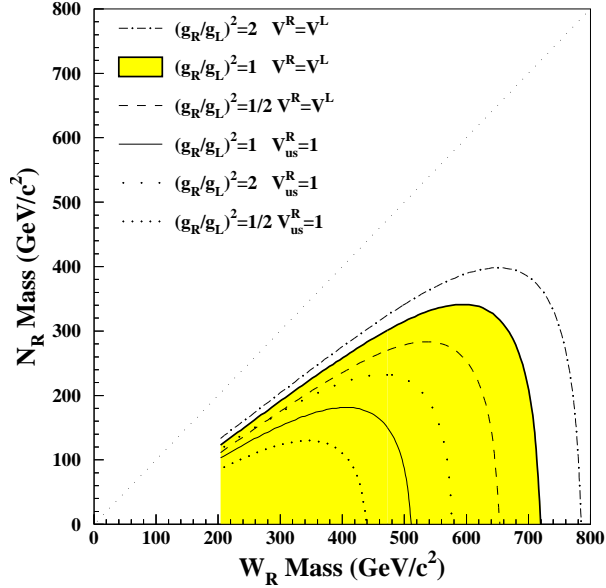}}
\vspace{0.3cm}
\caption{95\% CL excluded $W_R$ mass region from the {\sl peak search}.
The lines represent the contours for different values of the LRM parameters.
The diagonal line is the kinematic limit for the $W_R \rightarrow eN_R$
decay.}
\label{PRL_CONTOURS_1E}
\end{figure}

\indent
The resulting background subtracted upper limit,
including the effect of systematic uncertainties
(dominated by a 7.6\% uncertainty in the $W/Z$ background normalization),
is shown in
Fig.~\ref{PRL_LIMIT_SB_NEW}. Also shown is a second order ($\alpha_S^2$)
theoretical calculation \cite{Neerven} of $\sigma B$ assuming
$g_R=g_L$ and $V^R=V^L$. The next to leading order MRS(H) \cite{MRSH} parton
distributions were used for the calculation.
The branching fraction $B(W_R \rightarrow eN_R)$
was calculated taking into account the $N_R$ and $t$-quark masses and assuming
$m_{N_R^e}=m_{N_R^\mu}=m_{N_R^\tau}$.
For small $N_R$ mass, this fraction approaches the naive $\frac{1}{12}$ value.
Figure \ref{PRL_CONTOURS_1E} shows the corresponding excluded mass region.
The contours are shown for different values of the LRM parameters
$g_R$ and $V^R$ \cite{RIZZO_Model_Dependences}.
The extreme effect of varying $V^R$ is illustrated by displaying
the contour for a mixing matrix
with $V^R_{us}=1$ (thus $V^R_{ud}=0$ for $V^R$ unitary),
suppressing the primary $ud \rightarrow W_R$ production mechanism.
Because the limit from this part of the search was extracted from the
inclusive $p_T^e$ distribution, without additional topological
requirements, it is valid irrespective of the specific decay
mechanism for the $N_R$ or the $W$ helicity.

\indent
For the {\sl eejj search}, events were selected using a trigger that required
two electromagnetic energy clusters, each with $E_T>20$ GeV. After event
reconstruction, 22 events had two good isolated electrons with $E_T>25$ GeV and
two or more jets with $E_T>25$ GeV within a pseudorapidity range
$|\eta_{e,j}|<2.5$.
Events consistent with $Z+$jets production
were  rejected by demanding that the
invariant mass of the two electrons $m_{ee}$ be
outside the range $70 \leq m_{ee} \leq 110$ GeV/c$^2$.
Two events remained in the sample and were
therefore considered $W_R$ candidates.

\begin{table}[t]
\caption{Background estimates and event yields
for the $eej$ and $eejj$ samples.}
\vskip 0.5cm
\begin{tabular}{c|c|c}
Background & \multicolumn{2}{c}{Event Yield for $79.0 \pm 4.3$ pb$^{-1}$} \\
Process & $eej$     \hspace{0.50cm}      &
  $eejj$  \hspace{0.50cm} \\ \hline
$Z,\gamma^*$ & 12.84 $\pm$ 2.31 \hspace{0.50cm} &
 1.26 $\pm$  0.34 \hspace{0.50cm} \\
$t\overline{t}$ & 0.61 $\pm$ 0.35 \hspace{0.50cm} &
0.43 $\pm$  0.16  \hspace{0.50cm} \\
$WW$ & 0.13 $\pm$ 0.02 \hspace{0.50cm} &
0.01 $\pm$ 0.01   \hspace{0.50cm} \\
QCD & 9.90 $\pm$ 4.01 \hspace{0.60cm} &
 1.38 $\pm$ 0.68   \hspace{0.50cm} \\ \hline
Total & 23.48 $\pm$ 4.64 \hspace{0.60cm} &
 3.08 $\pm$ 0.78  \hspace{0.50cm} \\ \hline
Observed & 23    \hspace{0.50cm}  & 2    \hspace{0.50cm} \\
\end{tabular}
\label{Background_Table}
\end{table}

\begin{figure}[t]
\epsfxsize=3.375in
\centerline{\leavevmode\epsffile[0 0 567 520]{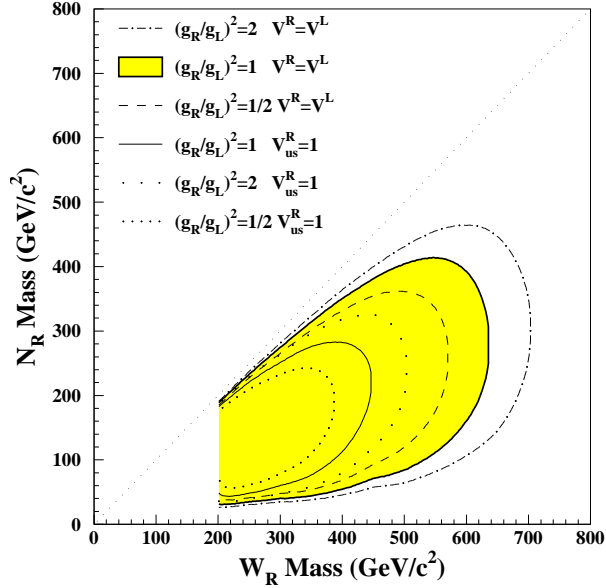}}
\vspace{0.3cm}
\caption{95\% CL excluded region of $W_R$ mass
from the {\sl eejj search} for the no mixing case.}
\label{PRL_CONTOURS_2E_NOMIX}
\end{figure}

\begin{figure}
\epsfxsize=3.375in
\centerline{\leavevmode\epsffile[0 0 567 520]{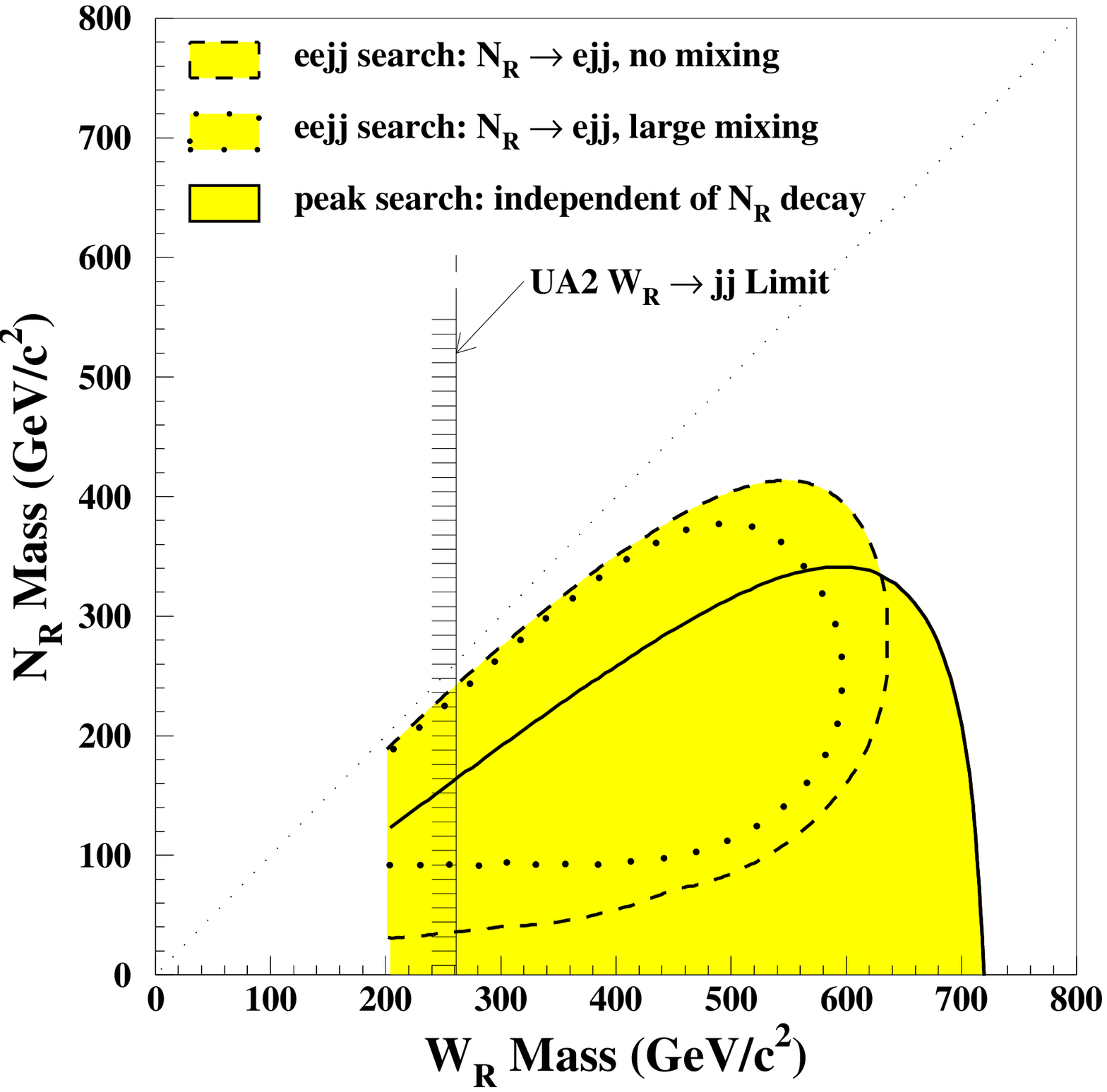}}
\vspace{0.3cm}
\caption{Excluded regions of $W_R$ mass at the 95\% CL assuming
$g_R=g_L$ and $V^R=V^L$ from the {\sl eejj} and {\sl peak searches}.}
\label{PRL_SUMMARY_CONTOURS_NEW}
\end{figure}

\indent
The largest background to the {\sl eejj} signal is multijet production (QCD)
with two jets misidentified as electrons.
To calculate this background, the invariant mass spectrum of
the jet pair with the largest electromagnetic fraction
in events with four or more jets was found.
This distribution was then scaled by a factor
determined from a two-component fit to the inclusive dielectron data using
the dielectron invariant mass spectrum from $Z,\gamma^*$ MC and the
measured inclusive dijet invariant mass spectrum.
The background from $Z,\gamma^*$+jets production was estimated by scaling the
number of observed events in the peak of the $m_{ee}$ distribution,
in events with two or more additional jets,  by
the tail-to-peak ratios obtained from MC.
Additional background is due to $t\overline{t}$ and $WW$ production.
The yield from $t\overline{t}$ was obtained using a Monte Carlo sample
with a detailed detector simulation and the
measured 6.4 $\pm$ 2.2 pb \cite{D0_top_observation} cross section.
For the $WW$ background, a sample of MC events and the theoretical
cross section were used.
To verify the background estimation, the yield of the above
processes to a final state with two electrons and one or more jets was also
calculated.
The background estimates and  event yields
are summarized in \mbox{Table \ref{Background_Table}} for the {\sl eej} and
{\sl eejj} final states.

\indent
As for the {\sl peak search}, the signal acceptance
for the {\sl eejj search} was calculated
for a grid of points in the
($m_{W_R}$,$m_{N_R}$) plane using MC simulation.
 The electron identification efficiency was
determined from $Z \rightarrow ee$ data. Example signal
efficiencies for the no mixing case are  (15.0 $\pm$ 1.7),
(10.1 $\pm$ 1.4) and (1.0 $\pm$ 0.4)\%
for ($m_{W_R}$,$m_{N_R}$) = (650,200), (400,350) and
(400,50) GeV/c$^2$, respectively.
 For the large mixing case the corresponding
efficiencies are lower due to the smaller $N_R \rightarrow eqq$ branching
fraction.
Also, for the large mixing case the search was restricted to
$m_{N_R} \geq 90$ GeV/c$^2$ since the efficiencies vanish when
$m_{N_R} \approx m_{W}$  due to a threshold effect.

\indent
Given no observed excess of events beyond the expected background,
we set a 95\% CL upper limit on $\sigma B$ using a Bayesian approach
\cite{PDG_Poisson} with a flat prior distribution for the signal cross section.
The uncertainties on the overall efficiency (10--20\%), the
integrated luminosity (5.5\%), and the background estimation (25\%)
were included in the limit calculation with Gaussian prior distributions.
The  resulting background subtracted upper limit is plotted in
Fig. \ref{PRL_LIMIT_SB_NEW},
while Fig. \ref{PRL_CONTOURS_2E_NOMIX} shows the excluded region
of the ($m_{W_R}$,$m_{N_R}$) plane for the no mixing case.

\indent
In conclusion, no evidence for the production of right-handed $W$ bosons
was found.
{}From a {\sl peak search} we set mass limits independent of the $N_R$ decay:
$m_{W_R}>650$ GeV/c$^2$ and $m_{W_R}>720$ GeV/c$^2$ at the 95\% CL,
valid for $m_{N_R}<\frac{1}{2}m_{W_R}$ and $m_{N_R} \ll m_{W_R}$ respectively,
assuming SM coupling ($g_R=g_L$ and $V^R=V^L$).
Also from the {\sl peak search}, we set a mass limit of $m_{W^\prime} > 720$
GeV/c$^2$ at the 95\% CL, extending the previous most stringent limit for heavy
left-handed $W$ bosons \cite{CDF_Wprime} which decay
into $e\nu$.
In addition, limits on $m_{W_R}$ valid for larger values of the $N_R$ mass
were obtained assuming that the $N_R$ decays to an electron and two jets.
Figure \ref{PRL_SUMMARY_CONTOURS_NEW} summarizes the results of the two methods
used for the search as an exclusion region in the ($m_{W_R}$,$m_{N_R}$) plane.
These limits on $m_{W_R}$ place stringent, though model dependent, limits on
possible $V+A$ couplings.

\indent
%
We thank the Fermilab Accelerator, Computing, and Research Divisions, and
the support staffs at the collaborating institutions for their contributions
to the success of this work.   We also acknowledge the support of
the U.S. Department of Energy,
the U.S. National Science Foundation,
the Commissariat \`a L'Energie Atomique in France,
the Ministry for Atomic Energy and the Ministry of Science and Technology
   Policy in Russia,
CNPq in Brazil,
the Departments of Atomic Energy and Science and Education in India,
Colciencias in Colombia,
CONACyT in Mexico,
the Ministry of Education, Research Foundation and KOSEF in Korea,
CONICET and UBACYT in Argentina,
and the A.P. Sloan Foundation.

\end{document}